\documentstyle[12pt]{article}
\begin{document}
% definitions
\newcommand{\newc}{\newcommand}

\newc{\be}{\begin{equation}}
\newc{\ee}{\end{equation}}
\newc{\ba}{\begin{eqnarray}}
\newc{\ea}{\end{eqnarray}}
\newc{\ie}{{\it i.e.}}
\newc{\eg}{{\it eg.}}
\newc{\etc}{{\it etc.}}
\newc{\etal}{{\it et al.}}

\newc{\ra}{\rightarrow}
\newc{\lra}{\leftrightarrow}
\newc{\no}{Nielsen-Olesen }
\newc{\lsim}{\buildrel{<}\over{\sim}}
\newc{\gsim}{\buildrel{>}\over{\sim}}

\begin{titlepage}
\begin{center}
June 1993\hfill
BROWN-HET-911 \\
             \hfill
% hep-ph/xxxxxxx
\vskip 1in

{\large \bf
A Cosmic String Specific Signature on the Cosmic Microwave Background
}
\vskip .4in
{\large R. Moessner\footnote{
Department of Physics, Brown University, Providence, R.I. 02912.}
\hspace{.2cm} L. Perivolaropoulos\footnote{Division of Theoretical
Astrophysics,
Harvard-Smithsonian Center for Astrophysics,
60 Garden St.,
Cambridge, Mass. 02138.}\hspace{.1cm}$^1$  \hspace{.2cm} R. Brandenberger$^1$}
\end{center}

\vskip .2in
\begin{abstract}
\noindent
Using an analytical model for the string network we show that the kurtosis of
cosmic
microwave background  (CMB) temperature  gradient maps is a good statistic to
distinguish
between the cosmic string model and inflationary models of structure formation.
The difference between the stringy and inflationary value for the kurtosis is
inversely
proportional to the angular resolution and to the number of strings per Hubble
volume of the strings'
scaling solution. If strings are indeed responsible for CMB anisotropies then
experiments with
resolutions of a couple of arcminutes or smaller could determine it using this
statistic.

\end{abstract}

\end{titlepage}
%\title{A Cosmic String Specific Signature on the CMB}
%\author{R.Moessner, \and L. Perivolaropoulos, \and R.Brandenberger }
%\maketitle

\section{Introduction}
Presently there are two main models for the formation of large-scale structure.
In the one, quantum fluctuations originating in an inflationary epoch become
classical density perturbations (Guth \& Pi 1982;
              Hawking 1982;
              Starobinsky 1982;
              Bardeen, Steinhardt \& Turner 1983), which grow by gravitational
instability into
the structures seen today. In the other, topological defects arising in a
phase transition in the early Universe act as seeds for galaxy formation
(Kibble 1976; Vilenkin 1985).
Galaxy redshift surveys, measurements of peculiar velocities of galaxies and
of temperature anisotropies in the cosmic microwave background (CMB) are
being used to test these theories. But so far experimental data have not been
sufficient to
rule in favour of one model or the other, being in more or less good agreement
with both
(see e.g. Brandenberger 1992).

 Our goal is to find, using an analytical approach, a signature of cosmic
strings
on the CMB which is unique to them and not predicted in inflationary models.
Cosmic strings
are one-dimensional topological defects which form when a symmetry is
spontaneously broken to
a smaller subgroup such that the first homotopy group
of the vacuum manifold is nontrivial. They are very thin lines of trapped
energy with mass per
unit length $\mu$; the spacetime around them is flat, but with a deficit angle
of $\alpha =
8\pi$G$\mu$ (Vilenkin 1981).

Inflationary models predict a gaussian distribution for the CMB temperature
anisotropies
measured at each angular scale
 (Efstathiou 1989). Temperature anisotropies from cosmic strings have been
shown numerically by Gott et al. (1990) and analytically by Perivolaropoulos
(1993b)
to be also nearly normally distributed,  in
spite of the inherently non-gaussian nature of the effect of a single string:
it causes
linear steplike discontinuities in the microwave sky
(Kaiser \& Stebbins 1984) (for a study of the statistics of density
perturbations
due to point-like seeds see Scherrer \& Bertschinger (1991)). By the central
limit theorem the
combined effect of all strings results in a gaussian signal for a large number
of strings .
However, the probability distribution for temperature {\em{gradients}} at small
enough angular
scales does show a departure from normality  (Gott et al. 1990). We use a model
for the
string network to calculate the gradient probability distribution (actually the
moment generating
function which is equivalent to it) and show that its kurtosis is a good
statistic to discriminate
between strings and inflation in CMB measurements at angular scales below a few
arcminutes. The
model is also implemented as a Monte-Carlo simulation, which is used to obtain
actual CMB maps from
cosmic strings and check the analytical results.

The reason why the distribution function for temperature gradients is more
nongaussian
than that for the anisotropies is that a superposition of $\delta$-function
perturbations, the
gradients of step-functions, approaches the gaussian distribution much more
slowly than the
superposition of the step-like temperature perturbations themselves.

We are making use of an analytical model (see e.g. Perivolaropoulos (1993a,b))
for the temperature anisotropies caused by strings which the
photons encounter between the time of last scattering and today. This model was
applied to the study of  peculiar velocities by Vachaspati (1992) and
Perivolaropoulos \&
Vachaspati (1993)). It has been used
 to
calculate the power spectrum and the amplitude of CMB anisotropies, as well as
their probability
distribution and corresponding moments. The temperature fluctuations were found
(Perivolaropoulos
1993a) (see also Hara, M\"ah\"onen and Miyoshi (1993)) to be well approximated
by a scale-invariant Harrison-Zel'dovich spectrum normalized to a
value consistent with observations of large-scale structure
(Perivolaropoulos, Brandenberger \& Stebbins 1990) (for other studies of large
scale structure
formation by cosmic strings see Hara \& Miyoshi (1993); Stebbins et. al.
(1987); Vollick (1992)).
The distribution function of temperature anisotropies was found
(Perivolaropoulos 1993b) to be
approximated by a gaussian to an accuracy of better than one percent. These
studies were useful
tests of the consistency of the cosmic string model with existing CMB data, but
did not provide a
way to distinguish it from other consistent theories based on inflation.

An alternative to our approach is a full numerical simulation of the evolution
of the
string network, and a subsequent numerical evolution of the photons through it
(Bennett, Stebbins \& Bouchet 1992). An
advantage of our analytical approach is that the explicit dependence on string
parameters and
angular scale can be shown, and the statistical fluctuations around the
predicted values for actual
experiments, due to having only a finite number of measurements, can be
calculated.

\section{Theoretical Model and Experimental Setup}

According to the scaling solution for cosmic strings
(Bennett \& Bouchet 1988;
Allen \& Shellard 1990), there is a fixed number  M$\approx10$ of strings with
curvature radius of
the order of the horizon per Hubble  volume at any given time. Their
orientations and velocities
are uncorrelated over distances larger than the horizon, and they move with an
rms velocity of
$v_{s}=0.15c$. In our model the strings are approximated as straight, so that
the Kaiser-Stebbins
formula for the temperature anisotropy due to the deficit angle of spacetime
around the string can
be used  (Stebbins 1988):
\begin{equation}
  \delta T/T=\pm 4 \pi G \mu \mid \hat{k} \cdot (\gamma_{s} \vec{v_{s}} \times
\hat{e_{s}}) \mid
\end{equation}
  where $\hat{k}$ is the direction of observation, $\vec{v_{s}}$ is the
velocity of the string,
$\hat{e_{s}}$ its orientation, and
$\gamma_{s}=(1-v_{s}/c)^{-1/2}$. Photons passing in front of the string are
redshifted, while
those passing behind it are blueshifted. The deficit angle extends only out to
a distance of one
Hubble radius ($H^{-1}$) from the string; this is due to compensation
(Traschen, Turok  \& Brandenberger 1986;
Veeraraghavan \& Stebbins 1990;
              Magueijo 1992): background matter
redistributes itself when strings are formed so that the monopole and
dipole moments of the energy density prturbations are zero.
Therefore photons passing
a string at a distance larger than the horizon scale are not affected by it.
This cutoff is quite
sharp  because of gravitational shock waves caused by the compensating
underdensities (Magueijo
1992).

The time between now and last scattering is divided into N Hubble times $t_{i}$
with
$t_{i+1}=2t_{i}$, and $N=log_{2}( t_{0}/t_{ls}) \approx 15$ for a redshift of
last scattering
$z_{ls}=1000$. The apparent angular size of a Hubble volume at time $t_{i}$ is
given by
$\theta_{H_{i}}\sim z_{i}^{-1/2} \sim t_{i}^{1/3} $ for large redshifts in the
matter dominated
era and assuming $\Omega_0=1$. Therefore,
 $\theta_{H_{i+1}}=2^{1/3} \theta_{H_{i}}$, and $\theta_{H_{ls}}=z_{ls}^{-1/2}
rad = 1.8^{\circ}$.
The strings encountered by the photon at subsequent Hubble times are assumed to
be uncorrelated.
This is justified because after a time $t_{i}$ a photon has traversed a large
part of its original
Hubble volume, so that at $2 t_{i}$ its new Hubble volume is mostly new space.
Moreover a string
moves a distance $v_{s} t_{i} = 0.15 c t_{i}$ which is approximately equal to
the strings' mean
separation, so that it is likely to encounter another string and intercommute
with it. At each
$t_{i}$ we assume the M string segments (each of length $2 H_{i}^{-1}$ since
the strings'
orientations and velocities are uncorrelated over larger distances) per Hubble
volume to have
random positions, orientations and velocities.

The effect of all strings together is taken to be a superposition of the effect
of the individual long
strings. Finally, the effect of loops is considered unimportant compared to
that of long strings, and the initial temperature inhomogeneities at the
surface of last scattering are assumed to be
negligible compared to those induced by the string network at later times.

Neglecting the effect of loops can be justified as follows: The typical loop
size is about  $10^{-4}$ of the Hubble radius (Bennett \& Bouchet 1988 ; Allen
\& Shellard 1990). At recombination this subtends an angle of $0.7''$. Only at
a redshift
$z \sim 2$ the angle becomes equal to $18''$, the smallest scale which we
consider in this paper. Thus, the loops are too small for the Kaiser-Stebbins
effect to apply. Instead, loops contribute to $\delta T/T$ via the usual
Sachs-Wolfe integral (Sachs \& Wolfe 1967), and the result of this integral is
given by $\delta T/T= \frac{1}{3} \Phi$, where $\Phi$ denotes the relativistic
potential at $t_{\rm ls}$. Since on scales corresponding to a beam width of
b$\geq 18''$ many small loops contribute to $\Phi$, the combined effect can (by
the central limit theorem) be expressed as gaussian noise. We will estimate the
magnitude of this noise in Section 6 and demonstrate that it is indeed
negligible, in agreement with the conclusions of Stebbins (1993).

Individual cosmic strings produce linelike discontinuities and plateaux in the
microwave
background. These are features which are most easily seen in temperature
gradient maps. Experiments
could either make line surveys or map out the temperature over a certain region
of the sky, say a
square of size  $\theta^{\circ} \times \theta^{\circ}$. Small angular
resolutions b provide a
better chance of seeing these features. We will quantify this statement in the
following sections
by calculating the moment generating function for the  temperature difference
between neighbouring
areas that can be resolved  by the experiment.

Fig.1 shows the result of a simulation for the temperature anisotropies
expected in an
experiment with $b=80''$ and $\theta = 2.2^{\circ}$ (i.e.$100 \times 100$
pixels).
 At each Hubble time $t_{i}$ $n_{i}$ string segments of length $\theta_{H_{i}}$
were placed at
random over an area of size $(\theta + \theta_{H_{i}} )^{2}$ (see Fig.2). By
the scaling solution
\begin{equation}
n_{i}=M (\theta+ \theta_{H_{i}})^{2} / \theta_{H_{i}}^{2}
\end{equation}
Note that the discontinuity lines seen in the map are not all at the position
of the strings
themselves. Each string at each of the 15 Hubble times gives rise to five such
lines, one at its
position, two at a distance $H^{-1}$ parallel to it, and another two because it
has finite length.
Since strings are one-dimensional objects, a square region with sides equal to
one quarter of the
horizon is crossed by about $M/4$ strings.      By equation (1) the string
shown changes the
temperature of region A  by $\beta r$ , and that of region B by  $-\beta r$,
where
\begin{equation} \beta \equiv 4 \pi G \mu \gamma_{s} v_{s} \end{equation} and
$r=| \hat{k} \cdot
(\hat{v_{s}} \times \hat{e_{s}}) |$,  $\hat{v_{s}}=\vec{v_{s}} / v_{s} $. The
direction of
observation $\hat{k}$ is approximately constant over the whole observed area,
and
$\hat{e}=\hat{e_{s}} \times \hat{v_{s}}$ is a random unit vector since the
strings' orientations
and velocities are random, so that $r=| \hat{k} \cdot \hat{e} |$ is uniformly
distributed over the
interval [0,1]. Each beam is assigned the temperature averaged over the area
covered by it.

\section{Moment Generating Function for Temperature Gradients}

Let X be the random variable (RV) whose realizations are the possible values
for the
temperature differences between two neighbouring beams (pixels). Our goal is to
derive its
distribution function. We do so by calculating the moment generating function
$M_{X}(t)$ of X. The
moment generating function of a RV is an important concept in statistics, it is
defined by
$M_{X}(t)=<e^{\textstyle tX}>$, where $<>$ denotes the expectation value. The
jth moment
$\mu_{j}=<X^{j}>$ of X is obtained from $M_{X}(t)$ by differentiation:
\begin{equation}
   \mu_{j}=\left(\frac{d^{j}}{dt^{j}}\right)_{t=0} M_{X}(t)
\end{equation}
An inversion formula gives the (cumulative) distribution function $F_{X}(x)
\equiv P(X \leq x)$,
the probability that $X \leq x$
(Lukacs 1960):
\begin{equation}
   F_{X}(x+\delta) - F_{X}(x-\delta)=\lim_{T \rightarrow \infty} \frac {1}{\pi}
\int_{-T}^{T}  \frac{ \sin t\delta}{t} e^{\textstyle -itx} \phi_{X}(t) dt
\end{equation}
provided that $x+\delta$ and $x-\delta$ are continuity points of F.
$\phi_{X}(t)=M_{X}(it)$ is called the characteristic function of X.
Equation (5) is the general formula for a RV that need neither be absolutely
continuous nor
purely discrete. If it is continuous, it can be shown from (5)
that a probability density $p(x)=F'(x)$ exists and that it is the Fourier
transform of
$\phi_{X}(t)$.

In the remaining part of this section we will derive $M_{X}(t)$. The reader who
is not
interested in the details is advised to jump to equation (21) and continue
reading from there.

Since the temperature difference results from the superposition of the effects
of the individual
strings, we can decompose X into $X=\sum_{i=1}^{15} X^{i}$, where $X^{i}$ is
the RV for the
temperature difference due to all the strings at Hubble time $t_{i}$. In turn
we can write $X^{i}$
as $X^{i}= \sum_{j=1}^{n_{i}} X_{j}^{i}$, where $X_{j}^{i}$ is the RV for the
temperature
difference due to the jth string at time $t_{i}$.

Now the moment generating function has the nice property that
$M_{X+Y}(t)=M_{X}(t) M_{Y}(t)$ if
X and Y are independent RV. Since the $X_{j}^{i}$ are independent in our model,
$M_{X}(t)$ can be
written as \begin{equation}
   M_{X}(t)= \prod_{i=1}^{15} M_{X^{i}}(t) \; \; \; \;, \; \; \; \; \; \; \; \;
M_{X^{i}}(t)=\prod_{j=1}^{n_{i}} M_{X_{j}^{i}}(t)
\end{equation}
Therefore only $M_{X_{j}^{i}}(t)=<e^{\textstyle tX_{j}^{i}}>$ has to be
calculated,
which is done below by determining the possible values for $X_{j}^{i}$ together
with their
probabilities.

The two beams have the same temperature if they both lie outside the string's
region of
influence or both in region A or B (see Fig. 2). This is true because the
observation direction
$\hat{k}$ is almost exactly the same for both beams for $b \leq
\theta_{H_{ls}}$. We are only
interested in such small angular resolutions because the simulations have shown
that for larger
ones the sought-after non-gaussian signature disappears. Moreover we will find
from the analytical
result that the non-gaussian signature goes to zero as b tends to
$\theta_{H_{ls}}$.
 The beams have unequal temperatures if either the string itself or one of the
lines of
discontinuity at Hubble volume cutoff (see dotted lines in Fig.2) pass through
either or both of
the beams.

First consider the string itself. The probability for it to cross the right
beam (see Fig.3 and
Fig.2) is
\begin{equation}
p^{i} \approx \frac {\theta_{H_{i}}/b} {((\theta_{H_{i}}+\theta)/b)^{2}}
\end{equation}
for $b \ll \theta_{H_{i}}$.
If it crosses the right beam it has uniform probability to be a distance $n
\frac {b}{2} ,
n \in [0,1]$, away from its centre, i.e. to be tangent to a circle $C_{n}$ of
radius $n
\frac{b}{2}$ around the centre. If it is tangent to  $C_{n}$ to the right of
the beam's centre and
such that it does not cross the left beam (i.e. somewhere along the arc between
points A and B),
then the possible values for $X_{j}^{i}$ are (for r=1) \begin{equation}
 X_{j}^{i}=\pm \beta(n-1) \; \; \; \;
\end{equation}
each with probability $p^{i} \frac{p_{1}}{2}$, where $p_{1}$ is the probability
for a string which is tangent to $C_{n}$ to be so between A and B. To obtain
Eq.(8) we have
used that $b \ll \theta_{H_{i}}$ and approximated the circular beams by
squares:
\begin{equation}
X_{j}^{i}=\pm[\frac{1}{b^{2}}(b(\frac{b}{2}-n\frac{b}{2})\beta-b(\frac{b}{2}+
n\frac{b}{2})\beta) - (-\beta)]= \pm\beta(1-n)
\end{equation}
 From Fig.3 it can be seen that
\begin{equation}
p_{1}=\frac {\alpha_{1}}{\pi}=\frac{1}{\pi} \arccos(\frac{1-n}{2})
\end{equation}
Similarly one obtains
\begin{equation}
X_{j}^{i}=\pm\beta(n+1)
\end{equation}
if the string is tangent to $C_{n}$ to the left of the beam's centre such that
it does
not cross the left beam (i.e. between C and D), each with probability
$p^{i}\frac{p_{2}}{2}$, where
\begin{equation}
p_{2}=\frac{1}{\pi}\arccos(\frac{n+1}{2})
\end{equation}
Finally, if the string is tangent to any other point of the circle, it passes
through the
left beam, and it has uniform probability to do so at distances $m\frac{b}{2}$
away from that
beam's centre, $m\in[0,1]$. For a given m the possible values are
\begin{equation}
X_{j}^{i}=\pm \beta(n+m) \; \; \;{\rm and} \pm\beta(n-m)
\end{equation}
each with probability $p^{i}\frac{p_{3}}{4}$, and $p_{3}=1-(p_{1}+p_{2})$.
Since we could have
started with the other beam, all the probabilities given above for the values
of $X_{j}^{i}$ have
to be multiplied by 2.

For each of the other four lines of discontinuity the same results apply, but
with $\beta$
replaced by $\beta/2$. The probability for $X_{j}^{i}=0$ is
$1-2p_{i}(1+4) = 1-10 p^{i}$. Finally, r is uniformly distributed over the
interval [0,1],
so $X_{j}^{i} \rightarrow rX_{j}^{i}$.

Putting all of this together (and setting $\beta=1$) we arrive at
\begin{eqnarray}
\lefteqn{M_{X_{j}^{i}}(t) = } \nonumber \\
      &  & 1-10p^{i} + 2p^{i} [ \int_{0}^{1} dr \int_{0}^{1} dm \int_{0}^{1} dn
\frac{p_{3}}{4} (e^{(n+m)tr}+\nonumber  \\
      &  &  e^{-(n+m)tr}+e^{(n-m)tr}+e^{-(n-m)tr} + \nonumber \\
      &  &  4(e^{(n+m)tr/2}+e^{-(n+m)tr/2}+e^{(n-m)tr/2}+e^{-(n-m)tr/2}) ) +
\nonumber \\
      &  & \int_{0}^{1}dr \int_{0}^{1} dn \{\frac{p_{1}}{2}(e^{(n-1)tr}+
e^{(-(n-1)tr} +
\nonumber \\
      &  &    4(e^{(n-1)tr/2}+e^{-(n-1)tr/2})) + \nonumber \\
      &  &  \frac{p_{2}}{2}(e^{(n+1)tr}+e^{-(n+1)tr}+
4(e^{(n+1)tr/2}+e^{-(n+1)tr/2})) \}]
\end{eqnarray}
We approximate $p_{1}$,$p_{2}$ and $p_{3}$ as constant over the interval $0
\leq n \leq 1$.
For $p_{1}=p_{3}=0.4$ and $p_{2}=0.2$ the integrals can be done with the result
\begin{eqnarray}
M_{X_{j}^{i}}(t) & = & 2p^{i} \{0.4 \frac{1}{t^2}
(-\sinh^{2}(t)-16\sinh^{2}(\frac{t}{2})) +
\nonumber \\
              &   & \frac{1}{t} (0.6{\rm shi}(2t)+5{\rm shi}(t) +1.6{\rm
shi}(\frac{t}{2})) \} +
1-10p^{i}
\end{eqnarray}
where ${\rm shi}(t)=\int_{0}^{t} dx \frac{\sinh x}{x}$. The next step is to
calculate
\begin{equation}
M_{X^{i}}(t)=\prod_{j=1}^{n_{i}} M_{X_{j}^{i}}(t)=(M_{X_{j}^{i}}(t))^{n_{i}}
\end{equation}
since the $X_{j}^{i}$ are identically distributed. Define
\begin{equation}
\lambda^{i}\equiv p^{i}n_{i} \approx M\frac{b}{\theta_{H_{i}}}
\end{equation}
Then
\begin{equation}
M_{X^{i}}(t)=\left(1+\frac{2\lambda^{i}(f(t)-5)}{n_{i}}\right)^{n_{i}}
\longrightarrow e^{2 \lambda^{i}(f(t)-5)}
\end{equation}
for $n^{i}$ large, which is the case here (f(t) is the function inside the
curly brackets
of Eq.(15)). This approximation makes it possible to do the final sum over
Hubble times:
\begin{equation}
M_{X}(t)=\prod_{i=1}^{15}M_{X^{i}}(t)=e^{2 \sum_{i=1}^{15} \lambda^{i}(f(t)-5)}
\end{equation}
 Since
\begin{equation}
\sum_{i=1}^{15}\lambda^{i}=Mb \sum_{i=1}^{15} \frac{1}{\theta_{H_{ls}}
2^{(i-1)/3}} =4.7
\frac{Mb}{\theta_{H_{ls}}}\equiv 4.7 \lambda
\end{equation}
the final expression for the moment generating function is
\begin{eqnarray}
\lefteqn{ M_{X}(t) = }  \nonumber \\
      &  &  e^{\textstyle 9.4\lambda( 0.4 \frac{1}{t^2}(-\sinh^{2}(t)-
16\sinh^{2}(\frac{t}{2})))}   \nonumber \\
      &  & \times e^{\textstyle 9.4\lambda( \frac{1}{t}(0.6{\rm shi}(2t)+ 5{\rm
shi}(t)+ 1.6{\rm shi}
(\frac{t}{2}))-5)}
\end{eqnarray}

\section{Probability Distribution and Kurtosis}

Fig.4 shows a plot of frequencies of temperature differences for $b=18''$
obtained from
the moment generating function $M_{X}(t)$ via the inversion formula Eq.(5) and
from a simulation.
The two curves are in good agreement. A gaussian distribution is shown for
comparison.

The kurtosis of the distribution of temperature gradients is a good statistic
to distinguish
between the string model and inflationary models. The kurtosis, defined as
\begin{equation}
 k_{4} \equiv \frac{<(X-<X>)^{4}}{<(X-<X>)^{2}>^{2}}
\end{equation}
measures the peakedness of a distribution and how far out its tails extend.
But this is
how a gradient distribution caused by strings differs from a gaussian one
(which is predicted
by inflationary models): The sharp discontinuities along the location of the
string produce
a higher number of large gradients, while the plateaux give a higher number of
small gradients
than would be expected in the gaussian case.

For a RV with zero mean
\begin{equation}
 k_{4}= \frac{\mu_{4}}{\mu_{2}^{2}}
\end{equation}
where $\mu_{j}$ is the jth moment given by Eq.(4).
A Taylor expansion of $M_{X}(t) $ around $t=0$ gives
\begin{equation}
 k_{4}=3+\frac{0.14}{\lambda}
\end{equation}
(see Fig.5).
The gaussian value for the kurtosis is 3. Recall that
$\lambda=M\frac{b}{\theta_{H_{ls}}}$.
So the departure from the gaussian value is inversely proportional to angular
resolution and to the
parameter M describing the string network. This is the main result of our
paper.  For $z_{ls}=1000$
and $M=10$ we have  \begin{equation}
 k_{4}=3+\frac{1.5}{{\rm b} \; {\rm(in \;  arcminutes)}}
\end{equation}
This shows that for beam sizes of a few arcminutes or smaller the string model
predicts a
non-gaussian kurtosis, whereas for larger beam sizes values very close to the
gaussian one are
predicted; consequently the latter experiments cannot be used to discriminate
between strings and
inflation using this statistic. Experiments with small angular resolutions have
not yet detected
CMB anisotropies. In the next section we discuss some of them to see whether
they should in future
measure a nongaussian value for $k_{4}$ if the anisotropies are caused by
strings.

\section{Confrontation with Observation}

Experiments only make a finite number of measurements and therefore do not
measure the true
moments. In order to distinguish between different models the distribution of
the measured
quantities around the exact ones has to be determined in each model.

An experiment as described in Section 2 makes $n=(\frac{\theta}{b})^{2}$
measurements of
temperature differences  between  beams lying next to each other along a
particular direction. We
can interpret the ith measurement as one realization of the RV $X^{(i)}$, where
$X^{(i)}$ is the RV
X of the previous sections for the ith pair of beams. If we define
\begin{equation} Y^{i}=
\frac{(X^{i}-<X^{i}>)^{4}}{<(X^{i}-<X^{i}>)^{2}>^{2}} \;\;\;\;\; i=1, \cdots ,n
\end{equation}
then the $Y^{(i)}$ are identically distributed, $\frac{1}{n} \sum_{i=1}^{n}
Y^{(i)}$ is the
measured kurtosis $k_{4}^{m}$, $< Y^{(i)}>$ is the true kurtosis $k_{4}$ of the
underlying
distribution, and \begin{equation}
Var(Y^{(1)}) \equiv <(Y^{(1)})^{2}>- <Y^{(1)}>^{2} = k_{8}-k_{4}^{2}
\end{equation}
where $k_{8}=\frac{\mu_{8}}{\mu_{2}^{4}}$ is the normalized eighth moment.
 By the central limit theorem the RV
\begin{equation}
\frac{\sum_{i=1}^{n} (Y^{(i)}-< Y^{(i)}>)} {\sqrt{n} \sqrt{Var(Y^{(1)})}}
\end{equation}
is normally distributed with unit variance for large n
(Feller 1971), i.e.
\[ P \left( \frac{\sum_{i=1}^{n} (Y^{(i)}-< Y^{(i)}>)} {\sqrt{n}
\sqrt{Var(Y^{(1)})}} \;\;\;
< \epsilon \right) = \Phi(\epsilon) \;\;\;\ ,\;\;
\Phi(\epsilon)=\frac{1}{\sqrt{2\pi}}
\int_{-\infty}^{\epsilon} dx e^{-x^{2}/2} \]  \[ \Leftrightarrow \;\; P \left(
\frac{1}{n}
\sum_{i=1}^{n} Y^{(i)} - <Y^{(1)}> \;\;\;<\epsilon
\sqrt{\frac{Var(Y^{(1)})}{n}} \right ) =
\Phi(\epsilon) \]  \begin{equation} \Leftrightarrow \;\; P(k_{4}^{m}-k_{4} \;\;
< \epsilon \;) =
\Phi\left(\epsilon\sqrt{\frac{n}{Var(Y^{(1)})}}\right)   \end{equation}
{}From Eqs.(27) and (29) we can see that the measured kurtosis is normally
distributed around
the true value with variance $\sigma^{2}= \frac{k_{8}-k_{4}^{2}}{n}$. For a
gaussian distribution
$k_{8}^{g}=105$, while for a stringy distribution the Taylor expansion of
$M_{X}(t)$ to eighth
order gives \begin{equation}
 k_{8} = 105 + \frac{29}{\lambda} + \frac{1.8}{\lambda^{2}}+
\frac{0.017}{\lambda^{3}}
\end{equation}

Fig.6 shows the distribution of the measured kurtosis about the predicted
values  in the two
models for an experiment by Fomalont et al. (1990),
who map out two areas each of size
$7'\times7'$ with angular resolutions ranging from $10''$ to $80''$. For a
resolution of $18''$ a
clear nongaussian signal is predicted, while for $b=80''$ the number of
measurements is not
sufficient to discriminate between the two models. If n were larger however, it
would be possible
to do so at $80''$.

A second experiment we want to discuss is one proposed by the Cambridge group
which is to start construction in the second half of 1993
(Lasenby 1992). It intends to map out an
area of size $10^{\circ}\times10^{\circ}$ at resolutions ranging from $10'$ to
$2^{\circ}$, and it
was planned, among other things, to look for possible signatures from
topological defects. Fig.7
shows that for its highest resolution of $10'$ and $M=10$ it would not be able
to tell whether the
underlying distribution is  gaussian or due to strings. It would be nice if
this new experiment
could be designed also to make measurements at the scale of 1 arcminute, in
which case it would be
able to rule in favour of or against strings. Fig.7 also shows a curve for a
different value of the
string scaling solution parameter M. If M should indeed be as low as 3 then
strings would give a
nongaussian kurtosis even at $10'$.

Based on our statistic, the Owens Valley experiment
(Readhead et al. 1989) with a beam width of $108''$ could
detect a signal for cosmic strings if a map of contiguous patches were
produced. Also, Melchiorri
et al. (1993) are planning a 3 m telescope balloon experiment with a beam width
of $2'$ which
will have the potential of detecting non-gaussian signatures from cosmic
strings.

At $10'$ the nongaussian signature has all but disappeared, so that for
experiments done at degree
scale or at several degrees (e.g. COBE where $b \approx 7^{\circ}$ ), strings
do {\em{not}} predict
a nongaussian distribution of either temperature gradients or temperature
values.

\section{Effect of Noise}

Gaussian noise diminishes the nongaussian signature expected from strings
alone.   The moment
generating function for the temperature difference between two
neighbouring beams due to gaussian noise is
\begin{equation}
 M_{Y}(t)=e^{\textstyle \frac{s^{2}t^{2}}{2}}
\end{equation}
where $s^{2}$ has to be determined for each kind of noise. The stringy moment
generating
function with  $\beta=4 \pi G \mu \gamma_{s} v_{s}$ not
set equal to 1 is
\begin{eqnarray}
\lefteqn{ M_{X}(t) = }  \nonumber \\
    &  & e^{\textstyle 9.4\lambda( 0.4 \frac{1}{(t \beta)^2}(-\sinh^{2}(\beta
t)- 16\sinh^{2}
(\frac{\beta t}{2})))} \nonumber \\
    &  &\times e^{\textstyle 9.4\lambda(  \frac{1}{\beta t}(0.6{\rm shi}(2
\beta t)+ 5{\rm shi}(\beta
t) + 1.6{\rm shi}(\frac{\beta t}{2}))-5)}
\end{eqnarray}
The measured temperature difference is given by X+Y, and since X and Y are
independent,
$M_{X+Y}(t)=M_{X}(t) M_{Y}(t)$.
Without noise the second moment of the probability distribution for temperature
differences is
$\mu_{2}= 5.43 \lambda \beta^{2}$ and
the kurtosis is $k_{4}= 3+\frac{0.14}{\lambda}$. With gaussian noise
the second moment is
\begin{equation}
\mu_{2}^{\rm noise}= 5.43 \lambda \beta^{2} + s^{2}
\end{equation}
and the kurtosis is
\begin{equation}
 k_{4}^{\rm noise}= 3+\frac{0.14}{\lambda} (1 + \frac{s^{2}}{5.43 \lambda
\beta^{2}} )^{-2}
\end{equation}
If $\mu_{2} = s^{2}$ the nongaussian part of the kurtosis is reduced by a
factor of four.
  This noise can either be instrumental or come from density perturbations at
last scattering.
Instrumental noise is uncorrelated, so that $s^{2} = 2 \sigma^{2}$ if the noise
has variance
$\sigma^{2}$ at each pixel. Gaussian noise from last scattering is correlated,
and some cold dark
matter models predict the correlation function (Gott et al. 1990)
 \begin{equation} C(\theta) =
C(0)\frac {1}{1+ \frac {\theta^{2}}{2 \alpha ^{2}}}
\end{equation}
for the temperature anisotropies (RV T) for scales $\theta < 2^{ \circ}$, where
$\alpha = 8'$
is the coherence angle, and $C(0) = \sigma^{2}$ is the variance of T. We can
approximate the
temperature difference between two neighbouring beams by $Y = b T'$ and use the
fact that if a
gaussian RV T has the correlation function $C(\theta)$, then its derivative
$T'$ has a gaussian
distribution with variance $- C''(0)$
(Vanmarcke 1983).
Since $M_{bX}(t) =
M_{X}(bt)$ , $s^{2} = - b^{2} C''(0) = \frac {b^{2}}{\alpha^{2}} \sigma^{2} $.
A more careful
analysis shows that this approximation is good for $b < \alpha$. Fig.8 shows
the dependence of the
kurtosis on noise from density perturbations at last scattering, and Fig.9 for
instrumental noise.
For resolutions of a few arcminutes or below, noise from density perturbations
at last scattering
does not pose much of a problem because it is correlated, but instrumental
noise can wash out the
stringy signal if it is too large.

As an example, we can use the above results to estimate the effect of cosmic
string loops on the kurtosis of the CMB spatial gradient map. As discussed in
section 2, loops contribute as gaussian noise via the Sachs-Wolfe integral.

The Sachs-Wolfe integral due to cosmic string loops has been estimated in
Brandenberger and Turok (1986) (see also Traschen, Turok and Brandenberger
(1986)). In that work it was shown that the largest loops at $\rm t_{\rm ls}$
dominate the Sachs-Wolfe integral. The result for $\Delta T (\theta)$, the
temperature difference between two points separated by an angle $\theta$,
scales as $\alpha^{1/4}$, where $\alpha  \rm t$ is the radius of the largest
loops at time t. Taking $\alpha = 1$ and using G$\mu = 2 \cdot 10^{-6}$ gave
the value
\begin{equation}
  \left \langle \left ( \frac{\Delta T}{T} \right )^{2} \right \rangle^{1/2}
\sim 5 \cdot 10^{-5} \sin^{1/2} \frac{\theta}{2}
\;\;\;\;\;\;\;\;\;\;\;\;\;\;\;\; \theta < 0.5^{\circ}
\end{equation}
For $\theta = 18''$, corresponding to the beam width of the VLA experiment, and
with $\alpha = 10^{-2}$ we obtain a value of $1.5 \cdot 10^{-7}$. So the
gaussian RV Y for the temperature gradients from loops has variance
\begin{equation}
   s^{2} =  \left \langle \left ( \frac{\Delta T}{T} \right )^{2} \right
\rangle (\theta = b)
\end{equation}
{}From Fig.9 it follows that the effect of loops on the kurtosis is negligible.

\section{Discussion}

We have looked at a statistic which can distinguish strings from inflation and
which can also be
calculated analytically in a model for the string network, showing the
dependence on string
parameters. This is useful as a complementary approach to numerical simulations
which have to deal
with singularities in the evolution equations of the strings leading to some
uncertainty about the
results. The statistic investigated here might not be the most sensitive one,
we intend to look at
others in the future. The genus curve (Gott et al. 1990) for example, a
topological statistic, makes use of
the two-dimensional pattern of the anisotropies. A multifractal analysis
(Pompilio 1993) is sensitive to
higher moments of the temperature gradients.

\section{Acknowledgements}
\par
This work was supported in part by the Department of Energy under contract
DE-FG02-91ER 40688 -
Task A (R.B. and R.M.) and by a CfA Postdoctoral Fellowship (L.P.).

\newpage
\section{Figure Captions}

{\bf Figure 1 :} Simulated map of CMB temperature anisotropies from strings for
a resolution
of $80''$. The map has size $2.2^{\circ} \times 2.2^{\circ}$ ($100 \times 100$
pixels). There are M=10 strings per Hubble volume, and units are such that
$\beta = 4 \pi G \mu \gamma_{s} v_{s} = 1$. The colour scale ranges from
$\delta T/T$ = -15 to +15.
\vskip .5cm
{\bf Figure 2 :} Model for the effect of a single string on the CMB temperature
of a patch of
sky of size $\theta \times \theta$ mapped out by circular beams.
Strings located within the big square of size $(\theta + \theta_{H})^{2}$ can
affect the
temperature of the small patch.
\vskip .5cm
{\bf Figure 3 :} Calculation of temperature difference of two adjacent beams
due to one string.
\vskip .5cm
{\bf Figure 4 :} Frequencies of temperature differences X (in units where
$\beta=1$) for
$b=18''$ and a binsize of 0.2 for strings. The gaussian fit is chosen to have
the same variance as
the result for strings.     \vskip .5cm
{\bf Figure 5 :} Kurtosis of the distribution of temperature differences
between adjacent beams
expected for strings (with $M=10$) as a function of beam size b.
\vskip .5cm
{\bf Figure 6 :} Probability density functions of kurtosis of temperature
gradients measured in
the experiment by Fomalont et al. for an underlying stringy (for M = 10) and
gaussian distribution.
\vskip .5cm
{\bf Figure 7 :} Probability density of kurtosis measured in the experiment by
Lasenby at $10'$
for strings (for $M=10$ and $M=3$) and an underlying gaussian distribution.
\vskip .5cm
{\bf Figure 8 :} Dependence of kurtosis of temperature gradients on noise from
density
perturbations at last scattering giving rise to temperature anisotropies  with
variance
$\sigma^2$ at
each pixel. Here and in the next figure the values $M=10$, $G\mu=2 \times
10^{-6}$  and $\gamma_{s} v_{s}=0.15$ are used.
 \vskip .5cm
{\bf Figure 9 :} Dependence of kurtosis of temperature gradients on
instrumental noise
(with variance $\sigma^2$ at each pixel).

\newpage
\centerline{\bf References}
\vspace{0.8cm}
\noindent Allen B. \& Shellard E. P. S. 1990, Phys.Rev.Lett. {\bf 64}, 119.

\noindent Bardeen J., Steinhardt P. \& M.Turner M. 1983, Phys.Rev. {\bf D28},
              679.

\noindent Bennett D. \& Bouchet F. 1988, Phys.Rev.Lett. {\bf 60}, 257.

\noindent Bennett D., Stebbins A. \& Bouchet F. 1992, Ap.J.(Lett.) {\bf 399},
              L5.

\noindent Brandenberger R. \& Turok N. 1986, Phys.Rev. {\bf D33}, 2182

\noindent Brandenberger R. 1992, 'Topological Defect Models of Structure \\
    ..... Formation After the COBE Discovery of CMB Anisotropies', \\
     ..... Brown preprint BROWN-HET-881 (1992), to be publ. in proc. \\
                     ..... of the International School of Astrophysics
"D.Chalonge", \\
..... 6-13 Sept. 1992, Erice, Italy, ed. N.Sanchez \\
..... (World Scientific, Singapore, 1993).

\noindent Efstathiou G. 1989, in 'Physics of the Early Universe', SUSSP 36,
1989, \\
..... ed. J.Peacock, A.Heavens \& A.Davies (IOP Publ., Bristol, 1990).

\noindent Feller W. 1971, 'An Introduction to Probability Theory and its \\
..... Applications', Wiley, New York.

\noindent Fomalont E. et al. 1993, Ap.J. {\bf 404}, 8.

\noindent Gott J. et al. 1990, Ap.J. {\bf 352}, 1.

\noindent Guth A. \& Pi S. -Y. 1982, Phys.Rev.Lett.{\bf 49}, 110.

\noindent Hara T., M\"ah\"onen P. \& Miyoshi S. 1993, Phys. Rev. {\bf 47},
2297.

\noindent Hara T. \& Miyoshi S. 1993, Ap. J. {\bf 405}, 419.

\noindent Hawking S. 1982, Phys.Lett. {\bf 115B}, 295.

\noindent Kaiser N. \& Stebbins A. 1984, Nature {\bf 310}, 391.

\noindent Kibble T. W. B. 1976, J.Phys. {\bf A9}, 1387.

\noindent Lasenby A. 1992, 'Ground-Based Observations of the Cosmic Microwave
\\
     ..... Background', to be publ. in proc. of the International \\
     ..... School of Astrophysics "D. Chalonge", 6-13 Sept. 1992, \\
     ..... Erice, Italy, ed. N. Sanchez (World Scientific, \\
     ..... Singapore, 1993).

\noindent Lukacs E. 1960, 'Characteristic Functions', C.Griffin \& Company,
              London.

\noindent Magueijo J. 1992, Phys.Rev. {\bf D46}, 1368.

\noindent Melchiorri F. 1993, private communication.

\noindent Perivolaropoulos L. 1993a, Phys.Lett. {\bf B298}, 305.

\noindent Perivolaropoulos L. 1993b, 'On the Statistics of CMB Fluctuations \\
      ..... induced by Topological Defects',  Phys. Rev. D, Aug. 15 1993.

\noindent Perivolaropoulos L. \& Vachaspati T. 1993, `Peculiar Velocities \\
     ..... and Microwave Background Anisotropies from Cosmic Strings', \\
     ..... submitted to Ap. J. (1993).

\noindent Perivolaropoulos L., Brandenberger R. \& Stebbins A. 1990,
Phys.Rev.\\
      ..... {\bf D41}, 1764.

\noindent Pompilio M. -P. 1993, private communication.

\noindent Readhead A. et al. 1989, Ap.J. {\bf 346}, 566.

\noindent Sachs R. \& Wolfe A. 1967, Ap.J. {\bf 147}, 3

\noindent Scherrer R. \& Bertschinger E. 1991, Ap. J. {\bf 381}, 349.

\noindent Starobinsky A. 1982, Phys.Lett. {\bf 117B}, 175.

\noindent Stebbins A. et al. 1987. Ap. J. {\bf 322}, 1.

\noindent Stebbins A. 1988, Ap.J. {\bf 327}, 584.

\noindent Stebbins A. 1993, Ann. NY Acad. of Sci. {\bf 688}, 824

\noindent Traschen J., Turok N. \& Brandenberger R. 1986, Phys.Rev. {\bf D34},
              919.

\noindent Vachaspati T. 1992, Phys.Lett. {\bf B282}, 305.

\noindent Vanmarcke E. 1983, 'Random Fields', MIT Press, Cambridge, Mass.

\noindent Veeraraghavan S. \& Stebbins A. 1990, Ap.J. {\bf 365}, 37.

\noindent Vilenkin A. 1981, Phys.Rev. {\bf D23}, 852.

\noindent Vilenkin A. 1985, Phys.Rep. {\bf 121}, 263.

\noindent Vollick D. N. 1992, Ap. J. {\bf 397}, 14.

\end{document}